\documentclass[aps,prb,preprint]{revtex4}
\usepackage{graphicx}
\usepackage{amsmath}
\usepackage{array}

\begin{document}

\title{%
Diffusion at the liquid-vapor interface}
\author{Daniel Duque}
\email{daniel.duque@uam.es}
\homepage{http://www.uam.es/daniel.duque}
\author{Pedro Tarazona}
\affiliation{%
Departamento de F\'{\i}sica Te\'orica de la Materia Condensada
and Instituto Nicol\'as Cabrera,
Facultad de Ciencias,
Universidad Aut\'onoma de Madrid,
Francisco Tom\'as y Valiente, 7. E-28049 Madrid, Spain.}
\author{Enrique Chac\'on}
\affiliation{%
Instituto de Ciencia de Materiales de Madrid, Consejo Superior
de Investigaciones Cient\'{\i}ficas,
E-28049 Madrid, Spain.}

\begin{abstract}
Recently, the intrinsic sampling method has been developed in order to
obtain, from molecular simulations, the intrinsic structure of the
liquid-vapor interface that is presupposed in the classical capillary
wave theory.  Our purpose here is to study dynamical processes at the
liquid-vapor interface, since this method allows tracking down and
analyzing the movement of \emph{surface molecules}, thus providing,
with great accuracy, dynamical information on molecules that are
``at'' the interface. We present results for the coefficients for
diffusion parallel and perpendicular to the liquid-vapor interface of
the Lennard-Jones fluid, as well as other time and length parameters
that characterize the diffusion process in this system. We also obtain
statistics of permanence and residence time. The generality of our
results is tested by varying the system size and the temperature; for
the later case, an existing model for alkali metals is also
considered. Our main conclusion is that, even if diffusion
coefficients can still be computed, the turnover processes, by which
molecules enter and leave the intrinsic surface, are as important as
diffusion. For example, the typical time required for a molecule to
traverse a molecular diameter is very similar to its residence time at
the surface.
\end{abstract}

\maketitle

\bibliographystyle{apsrev}

\section{Introduction}
\label{intro}

Inhomogeneous systems present a number of features that make them
intrinsically more complicated than bulk systems. The fact that the
equilibrium state of the system depends on the position causes a
number of physical quantities to be likewise dependent on the position
(such as the molecular number density), or even ill-defined (such as
the pressure tensor).  This also applies to dynamical properties ---
the most important one of these, (self-)diffusion, is complicated by
the fact that tracer molecules cannot be followed at will for any
given length of time, since they will enter and abandon zones that
have different dynamical properties.

It is therefore not possible in general to obtain
a value for the diffusion coefficient $D$ by the
well known Einstein relation for the
mean standard deviation (MSD) of the displacements
\begin{equation}
\label{eq:einstein1905}
\left<r^2\right> \xrightarrow[t\rightarrow\infty]{} 6 D t.
\end{equation}
The direct application of such a formalism to an inhomogeneous
fluid would result in an average result containing contributions
from the bulk phases and the interfaces. Indeed, $D$ can be
expected to have different values at different parts of the
system. (The same problem would arise of course
in the other main approach to $D$: by means of the Green-Kubo
formula involving the velocity autocorrelation function.)

This is true in particular for the best-known
%, and probably simpler,
inhomogeneous fluid system: the liquid-vapor interface.  In this
case, a liquid phase and its vapor are separated by an interfacial
region which, on average, is flat. The spacial dependence is therefore
limited in this case to one Cartesian coordinate, which we will take
as $z$.  For example, the mean density profile, $\rho(z)$, is obtained
from simulation by defining slab in the $z$ direction (``binning''),
and collecting occupation statistics for each of the slabs. This way,
a profile is obtained that shows two plateaus at constant values
corresponding to the liquid and vapor densities, and a typically
monotonic interfacial variation between them. Theoretical approaches,
from the pioneering van der Waals theory to the most recent density
functional approximations, may be used to directly obtain $\rho(z)$,
which depends only on the temperature, $T$, and on the molecular
interactions.  The density profiles may be much more structured in
other, apparently more complex, systems like a dense fluid against a
planar wall potential, but the apparent simplicity of the liquid-vapor
interface hides a much deeper difficulty.\cite{CWT,Evans} The
fluctuations of a free liquid surface have \emph{capillary wave} (CW)
modes with very low frequencies, and hence low excitation energies for
long wavelengths.  In the absence of any external potential, the
thermodynamic limit of a macroscopic free liquid surface becomes
undetermined, and the interface would be fully delocalized by the long
wavelength CWs. The Earth gravity field, which would be fully
irrelevant for the thermodynamic properties of one-phase systems up to
the scale of meters, becomes crucial to stabilize the liquid surface,
damping the CW fluctuations for wavelengths larger than millimeters,
and amplitudes larger than about one molecular diameter.  Still, the
mean density profile under Earth gravity conditions would be smoother
than the one observed with typical computer simulations, which employ
transverse box sizes in the range of $10-30$ molecular diameters (the
interfacial width can be estimated to be about twice as large in the
first case\cite{MolThCap}).  Within that limited range for the
transverse size the inclusion of the Earth gravity would be
irrelevant, but it is already possible to observe the dependence of
$\rho(z)$ with the transverse size of the simulation box.%
\cite{toxvaerd_1995,Sides_1999}

The CW fluctuations are a severe nuisance for the study of the
molecular diffusion at the liquid surface, since the molecular
kinetics, relevant for any physicochemical process at the surface, is
mixed with large collective fluctuations for the instantaneous
position of the surface, which give a smooth and size dependent
$\rho(z)$. The simplest estimations of surface diffusion properties
consider the molecules within thin slabs placed in the region with
inhomogeneous values of $\rho(z)$,\cite{townsend_1991,senapati_2002}
in some cases separating diffusion in parallel and perpendicular
components: Refs.~%
\onlinecite{meyer_1988,benjamin_1992,michael_1998,buhn_2004} that deal with
liquid-liquid interfaces, Refs.~%
\onlinecite{liu,chanda_2005,chanda_2005_2,chanda_2006,%
paul_2005,paul_2005_2,paul_2005_3,liu_2005,liu_2005,%
gonzalez_2006,clavero_2007} on liquid-vapor interfaces, and
Refs.~%
\onlinecite{lee_1994,lee_1994,martins_2004,bhide_2005,sega_2005,pal_2005,%
marti_2006,hijkoop_2007,thomas_2007}
on liquids (typically, water) adsorbed on, or confined in, different
substrates.
Another approach is to consider some operational definition of the
outmost liquid molecules and track their dynamics for a limited time.%
\cite{Taylor_1996,taylor_2003} Such procedures can only
yield a coarse distinction between bulk and surface properties, given
the obvious arbitrariness in the choice of parameters such as the
surface slab width, the definition of outmost molecules, the
complications associated with the diffusing molecules leaving and
entering the different domains, and the blurring effect of the
area-dependent CW fluctuations.

Classical \emph{capillary wave theory} (CWT)\cite{CWT} gives a
framework to interprete the effects of the CW fluctuations, and
provides an accurate extrapolation of the $\rho(z)$ profile obtained
in typical computer simulations to larger sampling sizes, including
the effects of weak gravity fields.  However, only over the last
decade has CWT become a practical tool to extract the \emph{intrinsic}
molecular properties of a liquid surface, from the broad distributions
produced by the CW fluctuations. The theory assumes that an
\emph{intrinsic surface} (IS) may be defined, to describe the
instantaneous boundary between the two coexisting phases, so that the
molecular distribution referred to that surface would give an
intrinsic density profile sharper that $\rho(z)$ and, more
importantly, independent of the transverse sampling size.  Over the
last decade, the increasing resolution in X-ray reflectivity data
allowed the deconvolution of the Gaussian CW distribution out of the
surface structure factor, and hence to obtain experimental results for
the intrinsic profile in cold liquid metal surfaces, with a clear
atomic layering structure.\cite{Persham}  Similar results were
obtained in computer simulations of simple fluid models,%
\cite{ourlayering,ourlayering2} whenever the frustration of the
freezing allowed to explore low temperatures, $T/T_\mathrm{c} < 0.2$,
relative to the critical one. As in experiments, the deconvolution is
possible because even $\rho(z)$ presents some layering at these low
temperatures.
These results indicate that the typical smooth shape of $\rho(z)$ in a
LV interface results from the convolution of the Gaussian CW
fluctuations with a strongly layered intrinsic profile, such that it
should be possible to identify, with reasonable confidence, the
\emph{outmost molecular layer} of the liquid phase.  More recently,
that concept led to the development of \emph{intrinsic sampling
method} in computer simulations,%
\cite{Chacon_Tarazona_PhysRevLett,Chacon_Tarazona_PhysRevB,IS} based
on the operational definition of the CWT intrinsic surface as a
geometrical locus for that first molecular layer, so that the
intrinsic profile, and other molecular intrinsic properties of the
liquid surface,\cite{ISM_aplications,ISM_aplications2} may be
extracted with accuracy and high resolution, without the CW blurring
observed in $\rho(z)$.  The intrinsic profile calculated with such
methods represent a direct, quantitative link between the generic
framework of the CWT and the computer simulations of liquid
surfaces. The result is a deeper structural understanding of the
surface: the capillary wave fluctuations, which cause an
area-dependent broadening of the mean profile $\rho(z)$, are absent in
the intrinsic one. These two profiles are, in general, remarkably
different: between the plateaus corresponding to the liquid and vapor
densities the intrinsic one is far from monotonic, showing a marked
layering. In fact, it much more resembles the pair correlation
function or the density profiles close to a hard wall.

The aim of our work here is to explore the application of the
intrinsic sampling method to the analysis of the molecular diffusion
at liquid surfaces. We may track down and analyze the movements of the
\emph{surface molecules}, i.e. those that define the IS. Thus we may
obtain, with great accuracy, dynamical information on molecules that
are ``at'' the interface, without the arbitrarity in the choice of a
surface slab, and making it possible to separate the molecular
diffusion on the surface from the fluctuations on the local position
of the surface.  The reader is referred to the previous references for
a complete description of the method, its variations, and their
results. The only relevant aspect here is that for each instantaneous
configuration of the system, this method selects a set of
\emph{surface molecules}, i.e. those identified as belonging to the
\emph{\it outmost liquid layer}, and called the IS \emph{pivots} in
the above references.  Within the intrinsic disorder of a liquid
surface we cannot expect that such molecular layer would have a sharp
definition --- it should indeed be regarded as a soft but robust
concept: changes in the set of parameters used for the operational IS
definition would produce small changes in the set of molecules which
are identifyed as belonging to the surface, but (at least for
$T/T_\mathrm{c}<0.8$) there is a clear optimal choice within rather
narrow windows for all parameters that must be fine tuned, as
commented at the end of the next section. In particular, the
two-dimensional density of the first liquid layer, i.e. the number of
surface molecules per unit area, is fairly well defined and provides
useful information on the molecular structure of a liquid surface,
which is blurred in the usual description in terms of $\rho(z)$.
\cite{our_new}
  
We begin with a section on methodology, Section \ref{methodology}, on
both the simulation details and the way the IS is calculated.  We then
consider diffusion in Section \ref{diffusion}, divided in three
subsections that go from the best known case, bulk diffusion, through
diffusion parallel to the interface, to the more involved case of
diffusion perpendicular to it.  We further analyze our results in
Section \ref{discussion}, where we consider
systems at other temperatures and other transverse
areas. We conclude with some remarks in \ref{conclusions}.

\section{methodology}
\label{methodology}

We consider the standard Lennard-Jones fluid,
in which molecules interact through a pairwise
potential of the form
\begin{equation}
\label{eq:LJ}
u(r)=
\epsilon
\left\{
  \left(
  \frac{\sigma}{r}
  \right)^{12}
-
  \left(
     \frac{\sigma}{r}
  \right)^6
-
  \left(
     \frac{\sigma}{r_\mathrm{c}}
  \right)^{12}
+
  \left(
     \frac{\sigma}{r_\mathrm{c}}
  \right)^6
\right\},
\end{equation}
 with interactions truncated at a cutoff radius of
$r_\mathrm{c}=3.02\sigma$.

In order to obtain dynamical information, molecular
dynamics (MD) simulations are performed,
using the software package \textsc{dl\_poly}.\cite{dlpoly}
The systems consists of a thick slab of liquid surrounded
by vapor. We therefore obtain two LV interfaces.
Since the two are independent (if the liquid is thick enough),
the properties measured in both should be averaged in order
to improve the accuracy, but for the sake of clarity we will
just present results obtained in one of the LV interfaces.

Our systems consist of $2592$ molecules, generally
(unless otherwise indicated)
at a temperature of
$k_\mathrm{B}T=0.678\epsilon$, which is our
estimate for the triple point temperature of the LJ fluid
(at this cutoff, see Ref.~\onlinecite{Matsy_de_Pablo} for
a discussion of the effect of truncation on this temperature).
The simulation cell is a square box of dimensions
$L \times L \times L_z$,
with (unless otherwise indicated)
$L=10.46\sigma$ and $L_z=90 \sigma$.
Periodic boundary
conditions are employed in all three directions.
The time step is set at a reduced value
of $dt=4.56\times 10^{-3} \sigma \sqrt{m/\epsilon}$.
The systems starts from either a crystalline configuration or a system
at other temperature and are equilibrated for $10^6$ time steps in the
$NVT$ ensemble (with a Nose-Hoover thermostat with a
time constant $10 dt$).

After this period, configurations are obtained
in the $NVE$ ensemble, in order to eliminate any
possible spurious effects of the thermostat on the
dynamics (in any case, we have checked these effects are
typically negligible). For the results
presented here, $50000$ configurations are
analyzed. We have found that only one configuration
out of ten needs to be analyzed, since the dynamics
is still slow for a time step of $10 dt$.
The analysis of the MSD is carried out in the standard
way; more sophisticated treatments aimed at reducing computer
storage\cite{understanding} are not needed in this case.

For each of these configurations, an IS analysis is carried out, as
described in Ref.~\onlinecite{IS}, requiring two operational parameters.
One of them, $n_\mathrm{s}$, has a very clear physical significance:
the two-dimensional density of the IS, i.e. the number of
\emph{surface molecules} to be selected per unit area.  Here we set
$n_\mathrm{s} \approx 0.8 \sigma^{-2}$, with a twofold motivation; on
one hand, this value had already been identified as the most physical
one from close inspection of the intrinsic profile for the LJ
fluid.\cite{IS}  On the other, the same set of simulations described
here can be used to independently obtain a value for this parameter,
confirming this value as the best one from a kinetic analysis (see
Ref.~\onlinecite{our_new} for details on this procedure).  The other
parameter, $q_\mathrm{u}$, fixes the maximum wave number to be used in
the Fourier representation of the IS. This mathematical surface
associated to each molecular configuration within the intrinsic
sampling method will not appear explicitly this article, but still it
is essential for the self-consistent procedure used to select the IS
molecules.  We have combined here a basis of $12$ planar waves for
each $x$ and $y$ direction, which with the transverse box size $L =
10.46 \sigma$ means a value of $q_\mathrm{u}=1.15 \times 2\pi/\sigma$,
within the optimal range of choices as explained in the above
references.

\begin{figure}
\includegraphics[width=8cm]{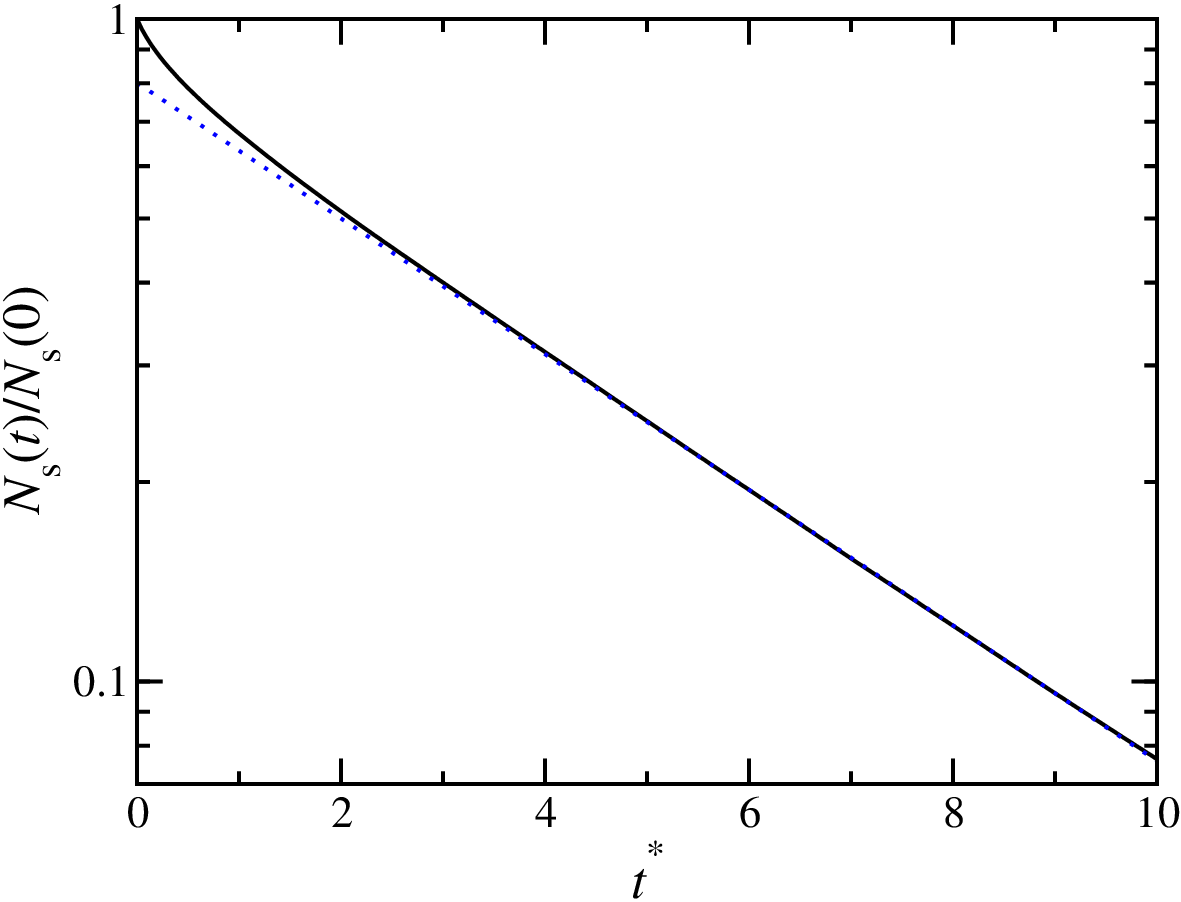}
\caption{\label{fig:diff2}
Mean fraction of molecules
which remain at the intrinsic surface for, at least, a time $t$,
$N_\mathrm{s}(t)/N_\mathrm{s}(0)$,
\textit{versus} reduced time $t$,
(units of $\sigma \sqrt{m/\epsilon}$).
Dotted line: linear regression of the
exponential decay at later times, shown to
intercept $t=0$ at about $0.8$.}
\end{figure}

In each analyzed configuration we select a number of surface
molecules, $N_\mathrm{s}=n_\mathrm{s} L^2$ (typically $88$ is our
optimal choice here) which self-consistently define the IS.  The
averages of the surface self-diffusion properties require us to follow
the individual history of each of these molecules, i.e. to identify
their permanence, exit, and possible reentrance in the IS list. All
the reported surface diffusion properties are associated to the
movement of molecules that stay continuously in the list of IS
molecules, and this requirement sets a clear limitation on the
avaliable sampling times. In Figure~\ref{fig:diff2}, $N_\mathrm{s}(t)$
represents the mean number of molecules which remain at the IS for, at
least, a time $t$; the exponential decay $N_\mathrm{s}(t) \sim
\exp(-t/\tau)$, with a typical decay time, or \emph{residence time} of
$\tau \approx 4.3 \sigma \sqrt{m/\epsilon}$ (at
$k_\mathrm{B}T=0.678\epsilon$), which we also list in
Table~\ref{table:diff1}.  This treatment mirrors the classical
definition of the bulk \textit{residence time} as the exponential
decay time for the process by which neighboring molecules drift apart.%
\cite{Impey_1983} It is also useful to keep in mind that for the case
of Argon with the usual LJ parameters $\epsilon/k_\mathrm{B}=119.8 K$
and $\sigma \approx 0.3405$nm one unit of reduced time corresponds to
$2.15$ picoseconds.  Thus, the decay time would be approximately
$9.2$ps for Argon.

This short value for the residence time implies that the sampling size
would rapidly decrease (and the statistical noise increase) for large
$t/\tau$.  The extrapolation to $t=0$ of this exponential decay at
longer times indicates that about $80\%$ of the molecules at the IS
will leave it at the constant rate of $1/\tau$, associated to the long
time exponential decay, while the remaining $20\%$ of the molecules
get out of the IS much faster, sometimes to undergo a rapid
reentrance. These molecules may be regarded as those which are only
marginally associated to the outmost liquid layer, e.g. those which
may be interpreted either as a local intrusion of the IS towards the
bulk liquid, or alternatively as a local extrusion of the next-outmost
layer towards the surface. The disordered structure of the liquid
makes the existence of such ambiguities unavoidable, and different
recipes to identify the IS from the molecular positions could make
different assignments to those molecules to be in, or out of, the list
of surface molecules. That is what makes the concept of the outmost
liquid layer a soft one, but at the same time a rather robust one,
since any reasonable choice of the tunable parameters would agree in
the selection of the large majority of the surface molecules. For the
diffusion properties analyzed here there is a further advantage, since
the relevant information to get the effective surface diffusion
coefficients comes from the sampling of molecular displacements at
the longest possible times, which automatically selects the properties
of those molecules with long permanence at the surface, representing
the outmost liquid layer with any sensible choice for the parameters
in the IS definition. The only practical inconvenience of the rapid
turnover of some surface molecules is that the time interval between
analyzed configurations, $\Delta t$, has to be short enough to make
sure that the estimation of $N_\mathrm{s}(t)$ is not affected by
further reduction --- in practice that is achieved, at
$k_\mathrm{B}T=0.678\epsilon$, with $\Delta t=10dt$, when only about
two and a half surface molecules are changed on average between
consecutive configurations.\cite{our_new}

\begin{figure}
\includegraphics[width=8cm]{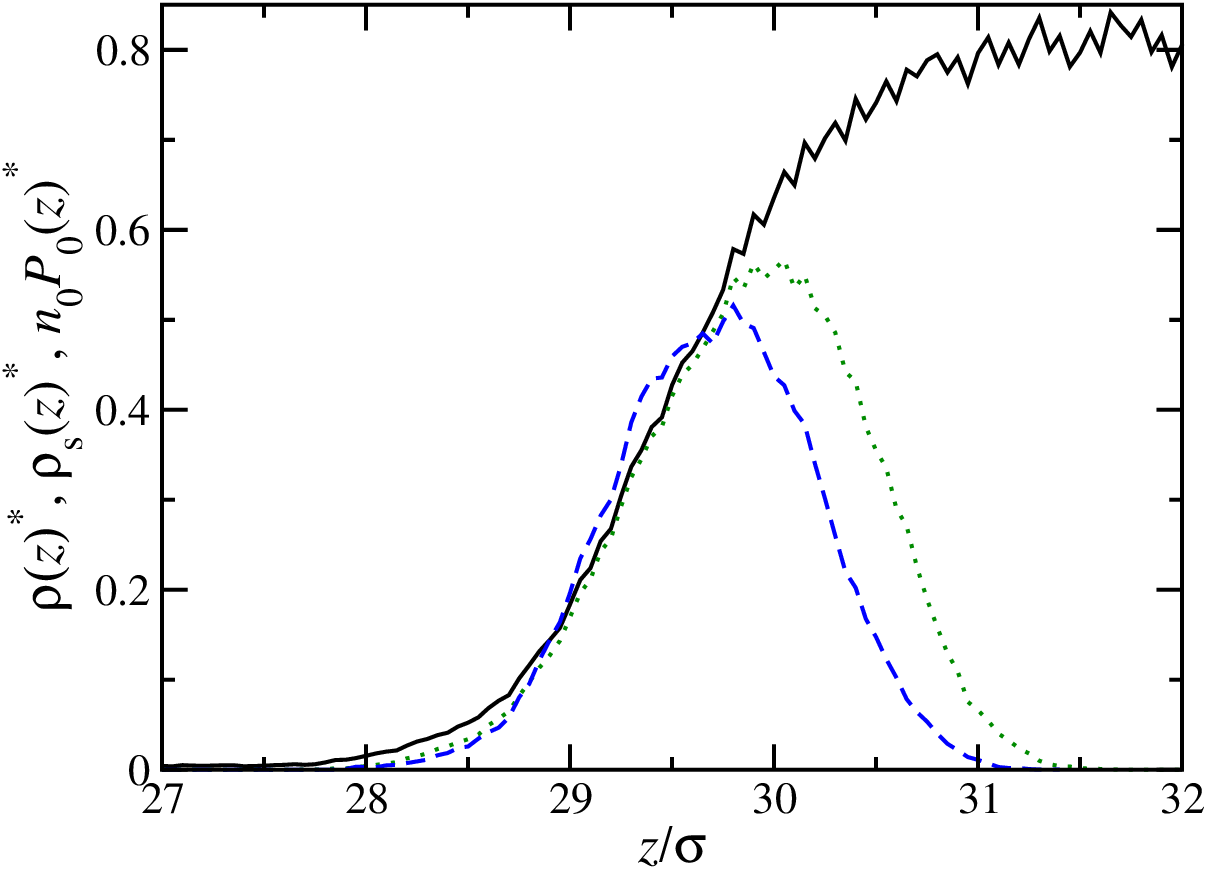}
\caption{\label{fig:rhos}
Mean density profile, %of the liquid-vapor interface,
$\rho(z)$ (solid line),
density profile of surface molecules,
$\rho_\mathrm{s}(z)$ (dotted line), and
density profile of old surface molecules,
$n_0 P_\mathrm{o}(z)$
(dashed line). The profiles are given in
reduced units of $\sigma^{-3}$.}
\end{figure}

The mean profile of the whole system, $\rho(z)$, and that restricted
to the surface molecules, $\rho_\mathrm{s}(z)$, are presented in
Figure~\ref{fig:rhos}, the later having a typical Gaussian shape which
represents the fluctuations of the IS, and which becomes wider with
increasing temperature and transverse sizes of the simulation
box. Notice that despite the mean profile character of
$\rho_\mathrm{s}(z)$, all the selected surface molecules lie exactly
on the instantaneous IS, we may therefore use them to follow exactly
the self-diffusion of the molecules at the outmost liquid layer,
without the coarsening effect of the CW fluctuations if we were
selecting the molecules within a fixed surface slab. Moreover, we may
keep track of the mean profiles for surface molecules which have
continuous permanence in the IS list for more than a time $t$, thus
defining a distribution $\rho_\mathrm{s}(z,t)$, whose integral over
$z$ is directly linked to the number of surface molecules older
than~$t$,
\begin{equation}
N_\mathrm{s}(t)= L^2 \int dz \rho_\mathrm{s}(z,t).
\end{equation}
We may expect that any diffusion property sampled for relatively large times would correspond to
molecules distributed as
\begin{equation}
\label{eq:factorizes}
\rho_\mathrm{s}(z,t)=n_0 P_\mathrm{o}(z) \exp(-t/\tau),
\end{equation}
where the time dependences factorizes into the same exponential decay
as $N_\mathrm{s}(t)$, and we define a $z$-distribution of \emph{old}
surface molecules, $P_\mathrm{o}(z)$, which is normalized to unity. The
prefactor has been discussed above to be $n_0\approx 0.8
n_\mathrm{s}$, i.e. representing about $80\%$ of the surface
molecules.  This is indeed the case, and in Figure~\ref{fig:rhos} we
compare the mean profile of the whole set of surface molecules,
$\rho_\mathrm{s}(z) \equiv \rho_\mathrm{s}(z,0)$, and the distribution
of old surface molecules normalized to $n_0$, i.e., $n_0 \
P_\mathrm{o}(z)$. The distribution of old surface molecules is
narrower, and asymmetric with respect to the whole one, and that may
be interpreted in terms of the rate at with molecules are incorporated
to, or deleted from, the list of surface molecules, as a function of
their position $z$.

\begin{figure}
\includegraphics[width=8cm]{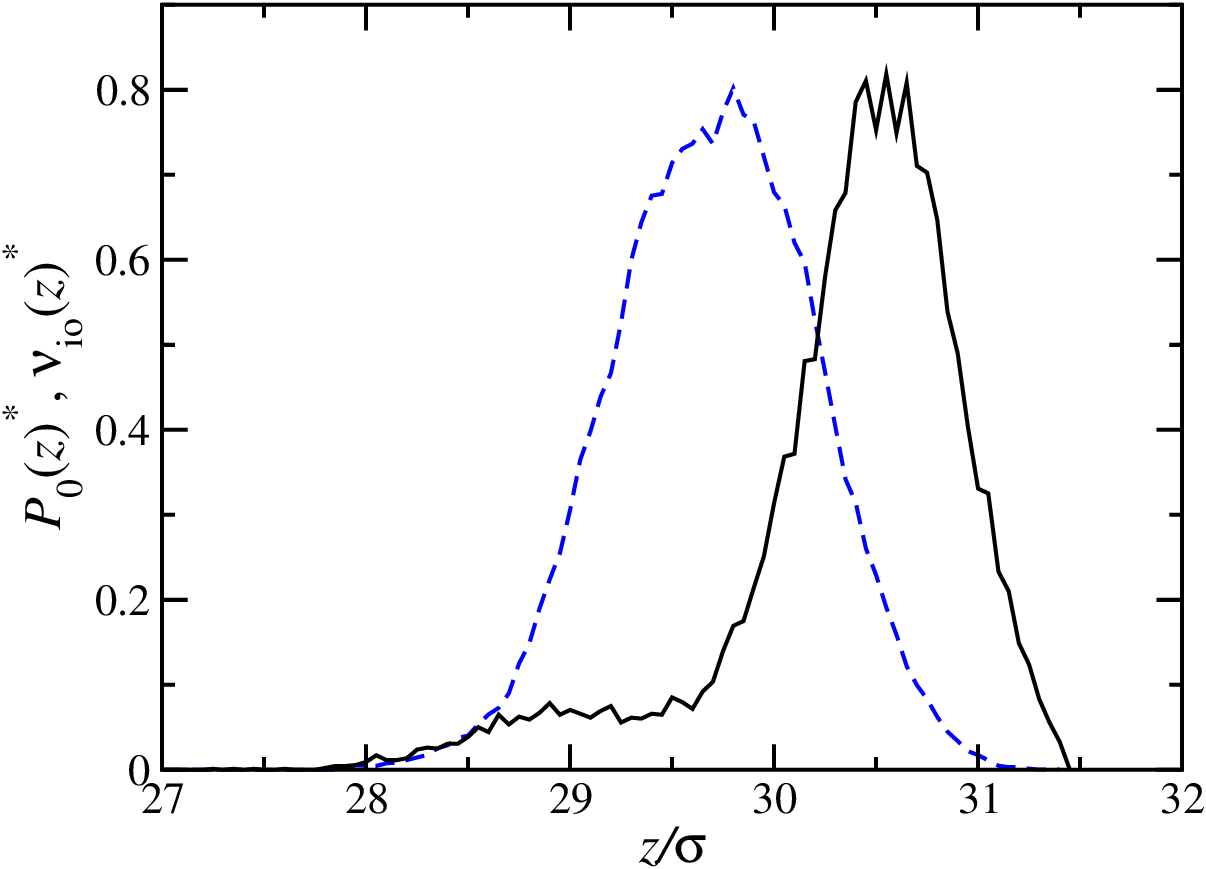}
\caption{\label{fig:distros}
Distribution of the input/output rate
per molecule, $\nu_\mathrm{io}(z)$, normalized to the
proper value set by Equation~\ref{eq:piv_t}
(hence, reduced in units of inverse time, $\sqrt{\epsilon/m}/\sigma$).
(solid line), together with $P_\mathrm{o}(z)$
in reduced units of $\sigma^{-1}$, normalized to unity (dashed line).}
\end{figure}

Since we select a fixed number of surface molecules in each
configuration, the loss of molecules from the IS layer, either towards
the liquid or the vapor sides, is always compensated by the
incorporation of new ones, and the time reversal symmetry of the MD
guarantees that the inflow and outflow of surface molecules at a given
value of $z$ are identical. We denote by $\nu_\mathrm{io}(z)$ that
input/output rate per molecule, which may be directly sampled along
our simulations and are presented in Figure~\ref{fig:distros},
together with $P_\mathrm{o}(z)$, for ease of comparison. The results
are fairly well compatible with rates being independent of the
previous permanence time of the molecule at the IS, so that
\begin{equation}
\frac{d}{dt}N_\mathrm{s}(t)= - L^2 \int dz \rho_\mathrm{s}(z,t) \nu_\mathrm{io}(z),
\end{equation}
which for large $t$ implies
\begin{equation}
\label{eq:piv_t}
\frac{1}{\tau}=-\frac{1}{N_\mathrm{s}(t)}\frac{d N_\mathrm{s}(t)}{dt}= \int dz P_\mathrm{o}(z) \nu_\mathrm{io}(z).
\end{equation}

The shape of $\nu_\mathrm{io}(z)$ is clearly composed of two
contributions: a large peak corresponding to the exit/entrance of
surface molecules to/from the bulk liquid, an a much smaller one
corresponding to the exit/entrance of surface molecules to/from the
bulk vapor.  In the following section we show how to extract the
surface diffusion coefficients from the information from the observed
displacements of the surface molecules, and the information on their
turnover distributions contained in $P_\mathrm{o}(z)$ and
$\nu_\mathrm{io}(z)$.

\section{Diffusion in the bulk, and at the intrinsic surface}
\label{diffusion}

We discuss diffusion for the particular choice of
temperature  $k_\mathrm{B}T=0.678\epsilon$, in three
different cases: bulk diffusion, diffusion parallel
to the interface, and diffusion perpendicular to it.

\begin{figure}
\includegraphics[width=8cm]{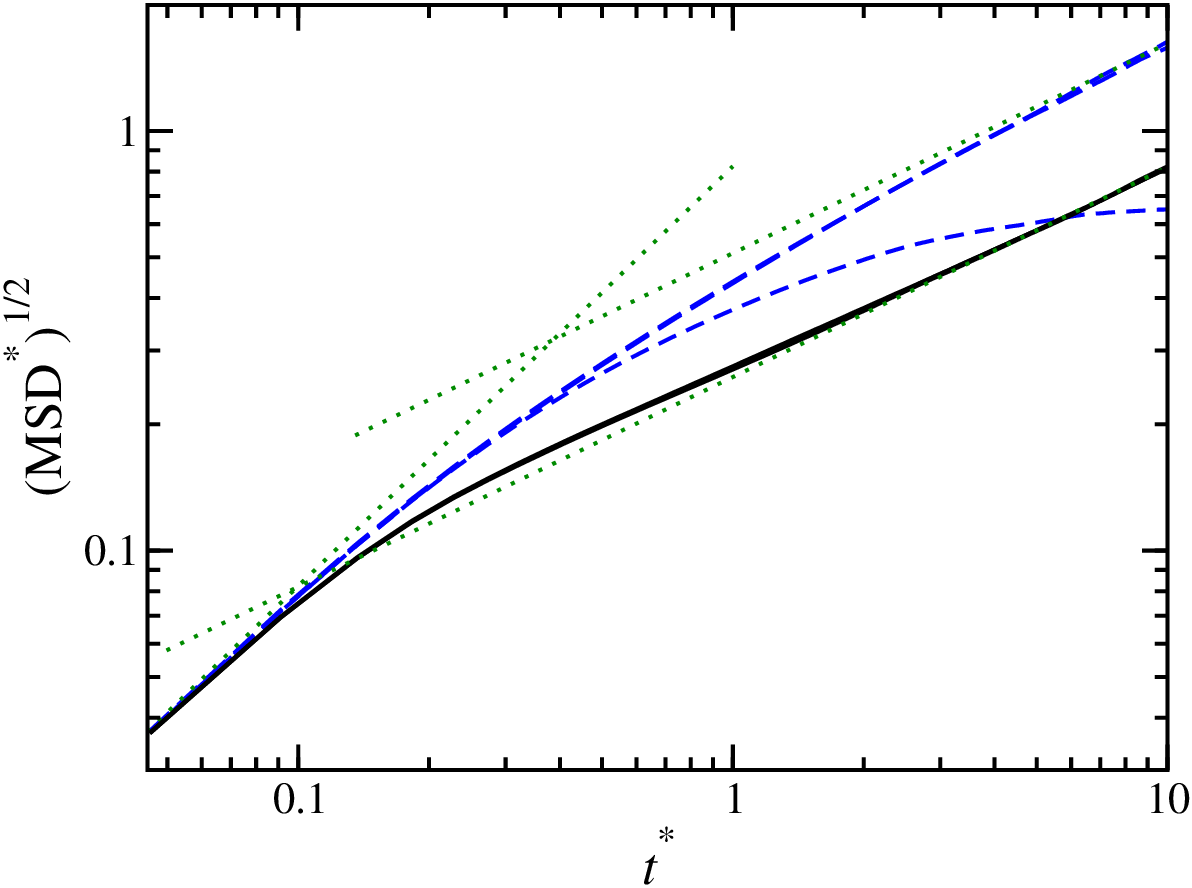}
\caption{\label{fig:diff1}
Square root of the mean standard deviation of displacements for
different subset of molecules \textit{versus} time, in reduced units:
$\sigma$ for the square root of MSDs and $\sigma \sqrt{m/\epsilon}$
for time.  Solid lines: $x$, $y$, and $z$ components in the bulk
liquid (hardly distinguishable), long-dashed lines: $x$ and $y$
components for the intrinsic surface (hardly distinguishable),
short-dashed line: $z$ component for the intrinsic surface. Dotted
lines: prediction from the Maxwellian velocity distribution,
Eq.~(\ref{eq:Maxwell}), and fits to Einsteinian diffusion equations,
Eq.~(\ref{eq:einstein_xyz}), with two values of the self diffusion
coefficient, one corresponding to the bulk liquid, the other
to the intrinsic surface.}
\end{figure}

\subsection{Diffusion in the bulk}
\label{bulk}

In Figure~\ref{fig:diff1} we plot the square root of the MSD for the
$x$, $y$, and $z$ components of the displacement
versus time for the bulk liquid phase.  
At short times all curves tend to the same line,
with a slope of $2$. This is the ballistic regime: times
so short that collisions can be neglected. 
In this case the Maxwellian distribution for velocities
directly provides a value for the MSD
\begin{equation}
\label{eq:Maxwell}
\left< x^2 \right> = \frac{k_\mathrm{B}T}{m} t^2 ,
\end{equation}
which is the dotted line in the graph.

At longer times a diffusive regime is reached,
with the Einstein relation for each of the
three Cartesian coordinates because of isotropy:
\begin{equation}
\label{eq:einstein_xyz}
\left<x^2\right> \rightarrow 2 D t; \qquad
\left<y^2\right> \rightarrow 2 D t; \qquad
\left<z^2\right> \rightarrow 2 D t.
\end{equation}

We indeed find no difference between the three components of the bulk.
From linear interpolation of this line in the $\log$-$\log$ graph one
reads $D_\mathrm{b}=0.034\sigma\sqrt{\epsilon/m}$, in good agreement
with previous data,\cite{meier,liu} see Table~\ref{table:diff1}.

The time for the crossover between the ballistic and the diffusive
regimes can be estimated from the crossing of the two linear
regression lines. In this case this is $t_\mathrm{c} \approx 0.099
\sigma \sqrt{m/\epsilon}$.
The corresponding mean displacement in each
direction would be $\Delta x_\mathrm{c}\approx 0.081\sigma$ if
read from the intercept of the lines, or $\approx 0.074\sigma$
from the MSD curves at $t_\mathrm{c}$. We
also list these numbers in Table~\ref{table:diff2}, where
we will also include results for other temperatures that
will be discussed in the next Section.
Since the density is $\rho_\mathrm{liq} =0.83 \sigma^{-3}$,
assuming a local coordination close to the FCC packing,
the typical intermolecular distance would be
$d=(\sqrt{2}/\rho)^{1/3} \approx 1.2\sigma$, and
collisions would take place with typical displacements
around $\Delta x \approx d-\sigma_0$, where $\sigma_0$ is
the minimum of the LJ potential, $2^{1/6}\sigma$.
This prediction yields $\Delta x \approx 0.072\sigma$,
consistent with the values found.

\newcolumntype{C}{>{$}c<{$}}

\begin{table}
\begin{tabular}{|c|c||C|C||C|C|C|C|C|}
\hline
 model& source & T^* &  \rho_\mathrm{liq}^*  & \tau^* & D_\mathrm{b}^* &  D_\|^*  &  D_\perp^* \\
\hline
 SA   & this work         & 0.212    & 1.17 & 13.2      & 0.020  & 0.038       & 0.04  \\
\hline
      & this work         &0.678   & 0.83 & 4.3         & 0.034  & 0.13         & 0.10  \\
 LJ   & Liu et al.\cite{liu} %
                          & 0.75^+ & 0.83 & \mathrm{-}  & 0.037  & 0.15         & 0.075 \\
      & Meier et al.\cite{meier} &0.678^* & 0.83 & \mathrm{-}  & 0.033  & \mathrm{-}   &  \mathrm{-} \\
\hline
 LJ   & this work         &0.848    & 0.74 & 2.27       & 0.082  & 0.20        & 0.16  \\
      & Meier et al.\cite{meier} &0.848^*  & 0.74 & \mathrm{-} & 0.078  & \mathrm{-}  & \mathrm{-}\\
\hline
\end{tabular}
\caption{\label{table:diff1}
Table of results, in reduced units.  Listed: type of model (SA: soft
alkali model of Ref.~\onlinecite{ourlayering}), source of the data quoted
((bulk values have been interpolated from data of Meier et al,
Ref.~\onlinecite{meier}), reduced temperature $T^* = k_\mathrm{B} T /
\epsilon $, reduced density of the liquid phase $\rho_\mathrm{liq}^*=
\rho_\mathrm{liq} \sigma^3 $, reduced residence time $\tau^*=\tau
\sqrt{\epsilon/m}/\sigma$, and reduced diffusion coefficients $D^*=D
\sqrt{m/\epsilon}/\sigma $: for the bulk ($ D_\mathrm{b}^*$), parallel
to the liquid-vapor interface ($D_\|^*$), and perpendicular to it
($D_\perp^*$).}
\end{table}

\subsection{Diffusion parallel to the interface}
\label{parallel}

Turning to the molecules at the interface, i.e. those selected as
\emph{surface molecules} by the intrinsic sampling method, their MDS,
along the three Cartesian coordinates also plotted in
Figure~\ref{fig:diff1}.  The ballistic regime at short times is the
same for all three components, and the same as for the bulk; but at
longer times the curves are very different from the bulk, the $x$ and
$y$, parallel, components remaining indistinguishable, the $z$
component differing. We focus on the parallel diffusion in this
section, for which we could expect the first two relations of
Eq. (\ref{eq:einstein_xyz}) to hold, but with a different value of the
diffusion coefficient:
\begin{equation}
\label{eq:einstein_xy}
\left<x^2\right> \rightarrow 2 D_\| t; \qquad
\left<y^2\right> \rightarrow 2 D_\| t.
\end{equation}
At long times, the MSD for the parallel components of our surface
molecules show this expected linear growth with $t$ limit, and we find
$D_{\|} = 0.13 \sigma\sqrt{\epsilon/m}$, Therefore, the parallel
diffusion coefficient is almost four times larger than the bulk one.
This remarkable difference, with similar factors, had already been
reported in works on water,\cite{townsend_1991,Taylor_1996,liu}
ethanol,\cite{taylor_2003} dimethyl sulfoxide,\cite{senapati_2002}
and liquid-liquid interfaces in LJ mixtures\cite{meyer_1988,buhn_2004}
(although not all of these works discriminate different components of
the diffusion coefficient) --- we should remark that, on the other
hand, very little change has been obtained for liquid-liquid
interfaces in mixtures of water and other polar liquids.%
\cite{benjamin_1992,michael_1998} Two works consider the liquid-vapor
interface of the LJ fluid: a value of $D_\| =0.13
\sigma\sqrt{\epsilon/m} $, at a similar temperature of
$k_\mathrm{B}T=0.75\epsilon$, is reported in Ref.~\onlinecite{liu} (also
included in Table~\ref{table:diff1} ), in good agreement with our
result. On the other hand, Ref.~\onlinecite{gonzalez_2006} find only a
twofold increase over the bulk diffusion coefficient, even if their
choice of temperature is again very close,
$k_\mathrm{B}T=0.75\epsilon$.

The time for the crossover between the ballistic and the diffusive
regimes is $t_\mathrm{c} \approx 0.38 \sigma\sqrt{m/\epsilon}$. The
corresponding mean displacement in each is either $\Delta
x_\mathrm{c}\approx 0.32\sigma$ (from the intercept of the lines), or
$\approx 0.23\sigma$ (from the MSD curves).  This means that the same
approximate increase with respect to the bulk by a factor of about three
applies to the characteristic time between collisions, 
the characteristic length between collisions, 
and the diffusion coefficient (this is of course consistent with
$\left<x^2\right> \propto D t$).

As shown in Figure~\ref{fig:diff1}, the crossover from the ballistic
to the diffusive behavior is quite different for the bulk and the
surface molecules. The MSD in the bulk converges to the diffusive
regime from above, i.e. after leaving the rapid ballistic regime,
there is a time interval in which the relative growth of the MSD is
slower than in the asymptotic diffusive regime. On the contrary, the
parallel diffusion of the surface molecules shows a much smoother
interpolation between the two limiting regimes, which may be
interpreted as a signature of the stronger disorder in the correlation
structure at the surface, causing a wider time distribution for
molecular rearrangement leading to the diffusive regime.  The MSD is
always below the asymptotic diffusion, and with very little difference
between the normal ($z$), and transverse ($x,y$) directions up to $t
\sim 0.4 \sigma\sqrt{m/\epsilon}$, and displacements $\sim 0.2 \sigma
$, which are approximately half way between the ballistic prediction
and the observed results for bulk for the same $t$.  These values are
also listed in Table~\ref{table:diff2}, under the heading of
``split.''

The surface molecules reach the diffusive regime only for typical
displacements larger than one molecular diameter $\sigma$, which
require a time similar to the residence time. If we compute the time
needed to diffuse to a displacement of $\sigma$ we find $t_\sigma =
4.2 \sigma \sqrt{m/\epsilon}$, a value very close to that of $\tau$
($4.3$ in reduced units).  This means that only about one third of the
molecules at the surface ($\approx 0.8\times e^{-1}$) remain in it
before diffusing a transverse distance similar to their diameter.  We
have to bear in mind the exponential decay for the number of molecules
which remain at the surface after a time $t$ --- e.g. a molecule would
likely move a distance $2 \sigma$ on the surface in a time $t \approx
4 \tau$, while in the bulk it would need typical times $10$ times
larger to diffuse the same distance. However, less than two percent of
the surface molecules ($\approx 0.8\times e^{-4}$) would remain as
such for $t=4 \tau$ and longer times, so that the large value of
$D_\|$ has limited relevance for the actual surface kinetics. The
turnover process of molecules from the surface to the bulk phases, and
its reverse, would be at least as relevant as the diffusion on the
surface. Therefore, it is most important to analyze that turnover
process, beyond its simple description in terms of the typical time
$\tau$. In the next subsection we show how the intrinsic sampling
method may also give an effective diffusion coefficient for the
movement of the surface molecules in the $z$ direction, in their
wandering which will eventually take them out of the surface.

%One could try to track down a correlation between residence
%time and diffusion. In order to do this, we calculate the
%MSD for pivots having different residence times...

\subsection{Diffusion perpendicular to the surface}
\label{perpendicular}

The naive relation
\begin{equation}
\label{eq:einstein_z}
\left<z^2\right> \rightarrow 2 D_\perp t
\end{equation}
is bound to fail if we restrict the averaging to the molecules at the
surface --- these are limited in space to the region about which the
IS fluctuates (both in position and in time).  The MSD for the $z$
coordinate will therefore tend to a constant value equal to the
squared width of the surface molecules density profile.  An
intermediate diffusive regime between ballistic behavior and this
final plateau may appear sometimes, but there is no reason to expect
this in general. Indeed, this is not the case in the situation
considered here, as is obvious from the dashed line in
Figure~\ref{fig:diff1}.

We show how to go beyond this naive prediction in a
series of steps. First, Einstein's equation is more
general than Eq.~(\ref{eq:einstein_z}): this would be
the second moment of a probability distribution:
\begin{equation}
P(z,t)=\frac{1}{\sqrt{4\pi D_\perp t}} \exp
   \left[ - \frac{z^2}{4 D_\perp t} \right].
\end{equation}
This, of course, is the solution to the
diffusion equation
\begin{equation}
\frac{\partial P(z,t)}{\partial t}=
D_\perp \frac{\partial^2 P(z,t)}{\partial z^2},
\end{equation}
with an initial Dirac delta function distribution: $ P(z,t=0)= 
%N_0
\delta(z) $, and vanishing values of $P$ and its flux for
large $\pm z$. In our MD simulations that $P(z,t)$ would correspond to
selecting molecules which at the time $t=0$ were within a very thin slab 
in the $z$ direction, and which in terms of their displacement $z$, 
would diffuse to have a probability distribution $P(z,t)$ after a time $t$. 

The same experiment could be done selecting the molecules on the
intrinsic surface at time $t=0$, and following their $z$ displacement
with $t$, but that would soon mix the surface and bulk diffusion as
commented in the introduction. Alternatively, we may represent in
$P(z,t)$ only the $z$ position of those molecules which have been
continuously at the IS from $t=0$.  That probability distribution
would not expand indefinitely, as required by the constant value of
the MSD for the $z$ component at long times in Figure~\ref{fig:diff1}.
That effect can be ascribed to an external potential $U(z)$ that
constrains surface molecules to a particular location in space.  Indeed, this
potential should be identified with a potential of mean force (PMF):
the mean-field description of the action of all other molecules on the
one diffusing on the surface.  Our diffusion equation would now be a
Smoluchowski equation:
\begin{equation}
\frac{\partial P(z,t)}{\partial t}=
D_\perp     \frac{\partial}{\partial z}
   \left[
    \frac{\partial P(z,t)}{\partial z}+
    \frac{1}{k_\mathrm{B} T}
         P(z,t) \frac{\partial U(z)}{\partial z}
  \right].
\end{equation}
(application of Smoluchowski equations to diffusion in non homogeneous
media can also be found in Refs.~\onlinecite{hijkoop_2007,liu}).  The
solution to this equation in the steady state would be an
\emph{equilibrium} distribution
\begin{equation}
P_\mathrm{eq}(z) \equiv
P(z,t\rightarrow\infty) \propto
 \exp
   \left[ - \frac{U(z)}{k_\mathrm{B} T} \right].
\end{equation}

A final ingredient comes from the fact that molecules
are continously leaving the IS,
a fact which may be incorporated through a loss
term in the Smoluchowski equation:
\begin{equation}
\label{eq:smoluloss}
\frac{\partial P(z,t)}{\partial t}=
D_\perp  \frac{\partial}{\partial z}
  \left[
    \frac{\partial P(z,t)}{\partial z}+
    \frac{1}{k_\mathrm{B} T}
         P(z,t) \frac{\partial U(z)}{\partial z}
  \right] - %\frac{1}{\tau} 
                \nu_\mathrm{io}(z) P(z,t),
\end{equation}
where we have to plug the input/output frequency at each $z$ position, $\nu_\mathrm{io}(z)$,
discussed above and shown in Figure~\ref{fig:distros}.
The function $P(z,t)$ solving this equation would not keep its normalization,
since its $z$ integral will decay with time, as the probability that an 
initially selected surface molecule would remain at the surface at time $t$.

For very long times we expect that, independently of the way we
initially select them, the remaining surface molecules will factorize
as in Equation (\ref{eq:factorizes}): $P(z,t) \sim \exp(-t/\tau)
P_\mathrm{o}(z)$, in terms of the probability distribution for the $z$
coordinates of old surface molecules, in Figure~\ref{fig:rhos}, and
decaying with the typical time $\tau$, so that (\ref{eq:smoluloss})
becomes
\begin{equation}
\label{eq:smoluloss2}
D_\perp \frac{d}{dz} 
 \left[
    \frac{d P_\mathrm{o}(z)}{dz} +
    \frac{P_\mathrm{o}(z)}{k_\mathrm{B} T} \frac{U(z)}{dz} +
 \right]
-\left[
\nu_\mathrm{io}(z) P_\mathrm{o}(z)-\frac{1}{\tau}
 \right] P_\mathrm{o}=0,
\end{equation}
which, for a given $D_\perp$ could be used to get the 
effective potential of mean force which is
compatible with forms of $P_\mathrm{o}(z)$ and $\nu_\mathrm{io}(z)$ 
obtained from our MD simulations, and related to the turnover 
time $\tau$ through (\ref{eq:piv_t}). 
The
potential of mean force could be easily obtained by integration of
Equation (\ref{eq:smoluloss2}):
\begin{equation}
\frac{1}{k_\mathrm{B} T}
\frac{dU(z)}{dz} =
-\frac{d (\log P_\mathrm{o})}{dz}+
 \frac{1}{ D_\perp }
   \int_{-\infty}^z dz'
       \left[
          \nu_\mathrm{io}(z') -
            \frac{1}{\tau}
       \right] P_\mathrm{o}(z').
\end{equation}
Thus, the only unknown in Equation (\ref{eq:smoluloss}) is the
parameter $D_\perp$, which is required both to get $U(z)$ and to set
the scale in the time variation of $P(z,t)$. If we integrate that
equation from the equilibrium probability distribution for all the
surface molecules, it would evolve in time towards the exponentially
decaying distribution of old surface molecules, and we could get
$D_\perp$ through the comparison of the actual evolution of $P(z,t)$
with the solution of the generalized Smoluchowski equation. A better
estimation of $D_\perp$ may be obtained if we start with a narrower
probability distribution $P(z,0)$, representing the surface molecules
which are initially within a thin slab around a given $z_0$. The
evolution of this probability distribution shows a first stage of
broadening, dominated by the diffusion so that $\left<(z-z_0)^2\right>
\rightarrow 2 D_\perp t$. In the later stage, the confining effects of
$U(z)$, and the losses regulated by $\nu_\mathrm{io}$ will saturate
the MDS and lead to the mature distribution $P(z,t) \sim \exp(-t/\tau)
P_\mathrm{o}(z)$, with a prefactor which would depend on the value of
$ D_\perp$.  The description of both the early and late stages with
the same effective diffusion coefficient, which should also be fairly
independent of the initial $z_0$ position of the chosen surface
molecules, is a strong requirement to confirm the validity of the
whole scheme, and hence the relevance of the best fitting value of
$D_\perp$ as a true normal surface diffusion coefficient.

\begin{figure}
\includegraphics[width=8cm]{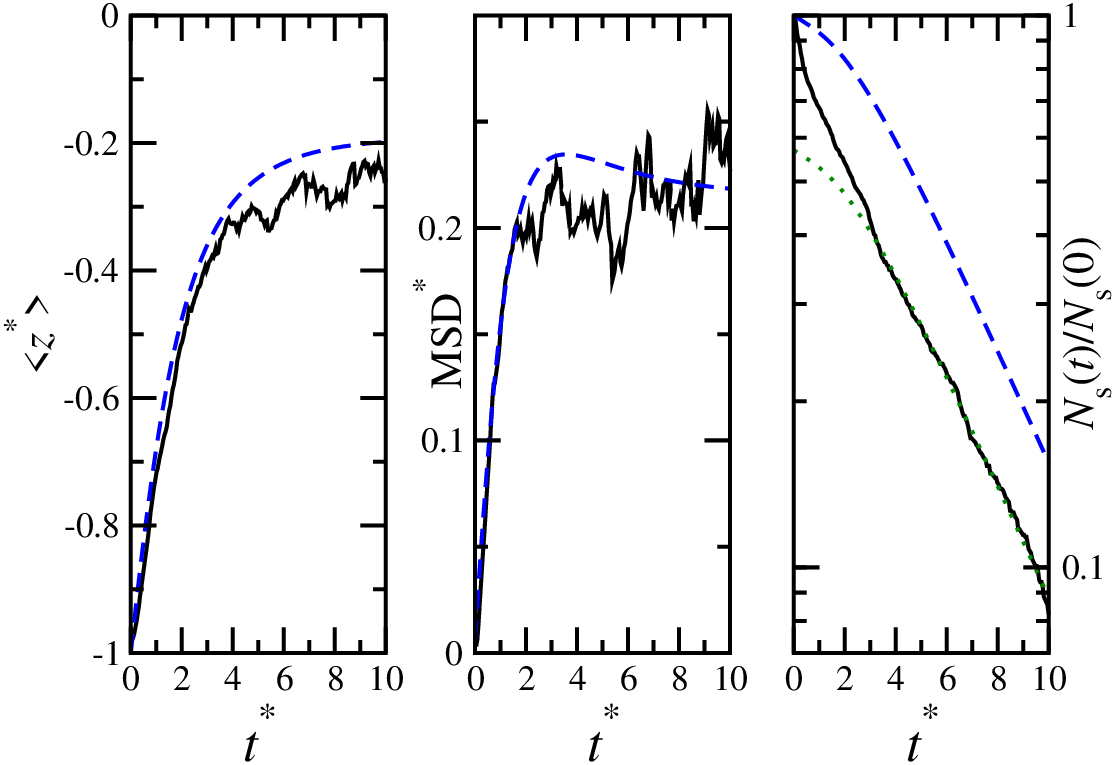}
\caption{\label{fig:drift}
Diffusion of a narrow peak, comparing
simulation results (solid lines) and predictions from
Eq.~(\ref{eq:smoluloss}) (dashed lines). Upper left: mean position
(first moment of the distributions), in units of $\sigma$, lower left:
MSD (second moment of the distributions), in units of $\sigma^2$,
right: fraction of surviving molecules
$N_\mathrm{s}(t)/N_\mathrm{s}(0)$ (normalization of the
distributions).  Dotted line in the later: rescaling of the dashed
line to demonstrate we obtain the proper residence time.  All graphs are
\textit{versus} reduced time $t$ (units of $\sigma
\sqrt{m/\epsilon}$).}
\end{figure}

In Figure \ref{fig:drift} we show results for the time evolution of a
peak initially at $z_0=-\sigma$ from the maximum of $P_\mathrm{o}(z)$
(its negative sign meaning toward the vapor side), with an initial
width of $0.2\sigma$.  The whole evolution of the $P(z,t)$ can be
traced, but we will just focus on its main features: its mean
position, its MSD, and its decay (normalization).  We try to fit these
three curves with the numerical solution of Eq.~(\ref{eq:smoluloss}),
and we obtain a very good agreement fitting a perpendicular diffusion
coefficient $D_\perp = (0.10\pm0.01)\sigma\sqrt{\epsilon/m}$ that is
comparable to, but smaller than the parallel one. Of course, the
exponential decay of $N(t)$ reproduces the previous value of
$\tau=4.26\sigma\sqrt{m/\epsilon}$ (see dotted line in
Fig.~\ref{fig:drift}). We have checked that our results are largely
independent of the initial position $z_0$, except for peaks very close
to the liquid phase. Indeed, the procedure seems to fail somewhat on
this side, presumably due to the more involved nature of the process
close to the liquid, with ``interstitial'' molecules, fluctuating
rapidly (ballistically) from being considered as belonging or not to
the IS. Those are again the expected limitations of any attempt to
locate the \emph{outmost liquid layer}, and we should be ready to
accept some degree of ``softness'' in that concept. Nevertheless, the
whole procedure provides a fairly well defined value of the normal
surface diffusion, with a value between those of the bulk and the
tangential surface diffusion, and close to the later.

Remarkably, there is only one work, to our knowledge, that discusses
normal diffusion in a liquid-vapor interface, Ref.~\onlinecite{liu}. They
report a value (also included in Table~\ref{table:diff1} ) that is
lower than ours, and closer to the bulk value. This is presumably due
to their use of slabs, by which some of slower bulk diffusion is mixed
with the normal diffusion. The remaining works all discuss
liquid-liquid interfaces, providing very similar values for both
diffusion coefficients at the interface,\cite{meyer_1988,buhn_2004} or
lower values for the normal coefficient%
\cite{benjamin_1992,michael_1998} (even lower, in fact, than the bulk
values). It is our feeling that this discrepancy stems mainly from the
saturation of the curve corresponding to normal diffusion that has
been discussed (short dashed curve in Figure \ref{fig:diff1}): if a
diffusion is computed for times shorter than the split time, a similar
value for both coefficients will be found; but for longer times, an
(effective) lower normal coefficient will be obtained.

\begin{table}
\begin{tabular}{|c|C|C||C|C|C|C|C|C|C|C|C|}
\hline
 & & & \multicolumn{3}{c|}{bulk} & \multicolumn{3}{c|}{IS} &
  \multicolumn{2}{c|}{split} &  \sigma  %  \multicolumn{1}{C|}{\sigma}  
 \\ \cline{4-12} 
 model & T^* &  \tau^* & 
  t_\mathrm{c} & \Delta x_\mathrm{c} \mathrm{(1)}  & \Delta x_\mathrm{c} \mathrm{(2)} &
  t_\mathrm{c} & \Delta x_\mathrm{c} \mathrm{(1)} & \Delta x_\mathrm{c}  \mathrm{(2)} &
  t_\mathrm{s} & \Delta x_\mathrm{s} &
  t_\sigma \\
      % & $D_\mathrm{b}^*$ &  $D_\|^* $ &  $D_\perp^*$ \\
\hline
SA & 0.212  & 13.2 & 0.19 & 0.087  & 0.078 & 0.36 & 0.16 & 0.14 & 0.8 & 0.2 & 13.2 \\
\hline
LJ & 0.678  & 4.3  & 0.099 & 0.081 & 0.074 & 0.38 & 0.32 & 0.23 & 0.4 & 0.2 & 4.2 \\ 
\hline
LJ & 0.848  & 2.27 & 0.19  & 0.18  & 0.14  &  0.47 & 0.43 & 0.32 & 0.3 & 0.2 & 2.9 \\
\hline
\end{tabular}
\caption{\label{table:diff2}
Table of relevant times and lengths for diffusion
at different temperatures, in reduced units:
$\sqrt{m/\epsilon}\sigma$ for time, $\sigma$
for lengths, $\epsilon/k_\mathrm{B}$
for temperature.
Listed: model type,
temperature, residence time (these first two
columns appear also in Table~\ref{table:diff1}),
crossover time and displacements for the bulk,
crossover time and displacements for the intrinsic surface,
split time and displacement,
time to cover a length of $\sigma$.
For the crossovers, two displacements are listed, one
from the intersection of the limiting lines, another
from the curves themselves. Details in the main text.}
\end{table}

\section{Discussion}
\label{discussion}

\begin{figure}
\includegraphics[width=8cm]{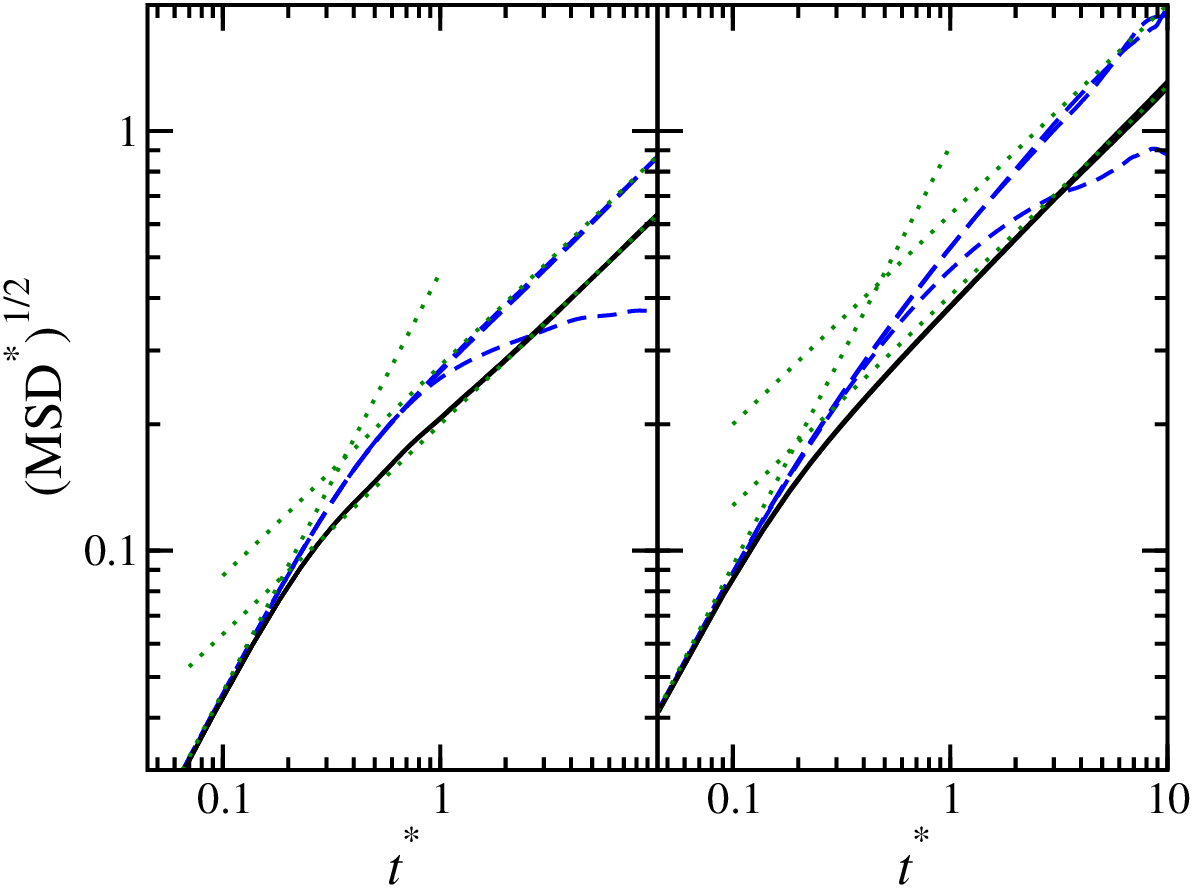}
\caption{\label{fig:otherTs}
Same as Figure~\ref{fig:diff1} for a lower temperature of
$k_\mathrm{B} T = 0.212\epsilon$, for the SA model (left), and a
higher one of $k_\mathrm{B} T = 0.848\epsilon$ for the same model, the
LJ fluid (right).  Solid lines: $x$, $y$, and $z$ components in the
bulk liquid (hardly distinguishable), long-dashed lines: $x$ and $y$
components for the intrinsic surface (hardly distinguishable, except
at the longest times due to statistical noise), short-dashed line: $z$
component for the intrinsic surface. Dotted lines: prediction from the
Maxwellian velocity distribution, Eq.~(\ref{eq:Maxwell}), and fits to
Einsteinian diffusion equations, Eq.~(\ref{eq:einstein_xyz}).}
\end{figure}

In order to support our main claims, we have performed simulations at
other temperatures. We would like to consider temperatures higher and
lower than the one considered. Hence, we have collected results for
the LJ liquid-vapor interface at a temperature $25\%$ higher,
$k_\mathrm{B}T=0.848\epsilon$.
It is not possible in principle to lower the temperature below the
triple point of LJ --- hence, in order to observe the influence of a
sizable drop in temperature, we have considered the soft alkali (SA)
potential of Ref.~\onlinecite{Chacon_Tarazona_PhysRevLett}.  The
parameters in this model have a similar meaning to the ones in the LJ:
$\sigma$ is the distance at which the potential vanishes and
$\epsilon$ is related to the volume integral of the intermolecular
potential function. Thus, the estimated critical temperature for this
model is $k_\mathrm{B}T_\mathrm{c}\approx 1.43\epsilon$, similar to
the one of the LJ fluid ($k_\mathrm{B}T_\mathrm{c}\approx
1.33\epsilon$).  This is a potential engineered to result in a very
low triple point temperature of $k_\mathrm{B}T_\mathrm{t}\approx
0.15\epsilon$.  We have considered a temperature of $k_\mathrm{B}T=
0.212\epsilon$.  The lateral size is $L=9.025\sigma$, and the number
of components employed in the IS Fourier analysis is $8$.  In
addition, the optimum value of the surface density is, for this model
$n_\mathrm{s}=0.7\sigma^{-2}$.  The curves showing the square root of
the MSD as a function of time, for the two temperatures, are given in
Figure \ref{fig:otherTs}.

The results for the residence time and for the diffusion coefficients
are included in Table~\ref{table:diff1}.  We see that with increasing
temperature decay diffusion becomes faster and residence times
shorter, as is to be expected. The ratio $D_\|/D_\mathrm{b}$ is about
$2.4$ for the higher temperature (LJ fluid), since the increase in
$D_\|$ is less drastic than that in $D_\mathrm{b}$. For the lower
temperature (SA fluid) this ratio is about $2$: in this case the
reduction in $D_\|$ turns out to be more drastic than that in
$D_\mathrm{b}$.  This is probably due to the fluid phase being already
very dense, while the molecules at the surface still have the option
of hopping along it --- surface diffusion is therefore reduced due to
the prevalence of energetic attraction (exerted mainly from the bulk
liquid) over entropy at lower temperatures.

We have also repeated the analysis of typical times and lengths for
the curves in Figure \ref{fig:otherTs}, and collected this information
in Table~\ref{table:diff2}.  The most obvious correlation is between
the time needed to diffuse a length of $\sigma$, $t_\sigma$ and the
residence time. In all three cases, these two quantities are nearly
equal. This corresponds to a natural assumption of a certain isotropy
in the diffusion process: by the time a molecule at the IS has
displaced a length of one molecular diameter, it is likely that the
molecule has also left the IS (which is of course consistent with the
IS being of molecular width). In other words, the product of $D_\|$
and $\tau$ should be nearly constant, which indeed is true, providing
a typical length of $\sqrt{D_\|\tau}\approx 0.7\sigma$ in all three
cases.

There seems to be some agreement in the split length, the approximate
displacement beyond which the parallel and perpendicular diffusion
curves split, with a value around $0.2\sigma$ in all cases.  Other
correlations are seen to be model-dependent. For example, the trend
for typical bulk crossover times and lengths differs in the SA model
due to its shallower potential well. Indeed, the argument given above
in terms of a local coordination close to the FCC packing is still
valid for the higher temperature, yielding a value of $\Delta x
\approx d-\sigma_0 = 0.12\sigma $ (to compare against $0.14\sigma$),
but fails for the SA model, with a value of $d$ that is below the
appropriate $\sigma_0\approx 1.5\sigma$.

In addition to varying the temperature, it is natural to explore the
dependence of diffusion with the interfacial area. On one hand, the CW
framework leads us to think that there could be a change of $D_\perp$
with the area. On the other hand, if this quantity is a true intrinsic
property of the surface, it should be area-independent.
In order to explore this possibility, we have scaled the lateral
length of our simulation box, $L$, by factors of $1/2$, $2/3$, $3/2$,
and $2$ (i.e., the area going from $1/4$ to $4$ times the original).
For the last two cases, we have increased the number of molecules to
$8000$ and $15625$, respectively (otherwise, the liquid film would
have been too thin); we have also enlarged the cell in the $z$
direction, with $L_z=200$ for the smallest area in order to
accommodate a thicker liquid slab.  In order to keep the same
resolution in the IS Fourier treatment, the $12$ functions for the
original area will now be $6$, $8$, $18$, and $24$, respectively.

\begin{figure}
\includegraphics[width=8cm]{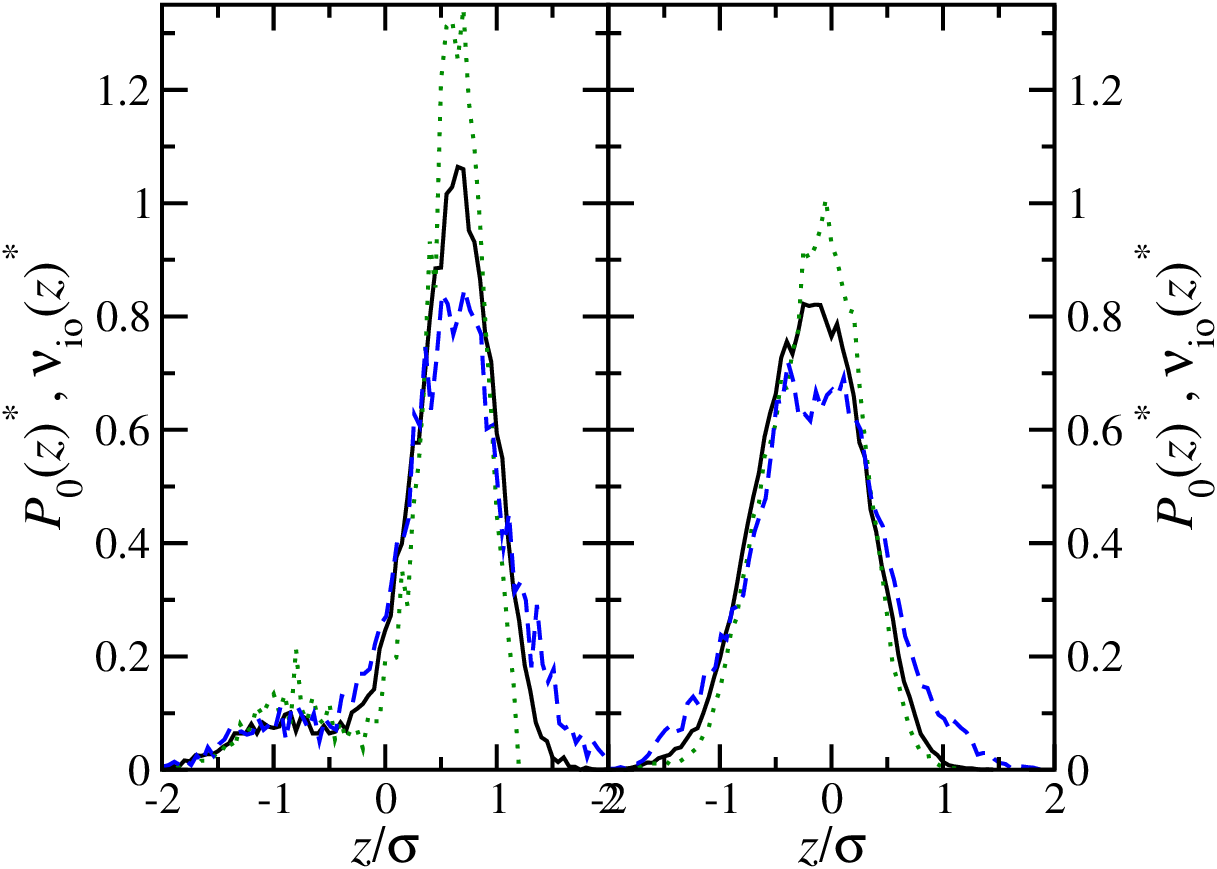}
\caption{\label{fig:distros2}
Comparison of the different distribution functions for different
lateral sizes.  Left: input/output rate per molecule,
$\nu_\mathrm{io}(z)$; right: density profile for old surface molecules,
$P_\mathrm{o}(z)$.  Solid lines: original lateral size $L=10.46\sigma$
(same curves as in Fig. \ref{fig:distros}); dashed line: lateral size
twice longer; dotted lines: lateral size twice shorter.  All functions
are normalized to unity.}
\end{figure}

As shown in Figure~\ref{fig:distros2}, functions $P_\mathrm{o}(z)$ and
$\nu_\mathrm{io}(z)$ contract for smaller areas and stretch for
larger ones, as is to be expected from the general broadening due to
CWs.  (In order to obtain these curves, it is important that the
various properties of the IS should be computed by subtracting the $z$
value of the $q=0$ (constant) mode --- otherwise, bulk density
fluctuations produce a spurious broadening of the distributions at
smaller areas,\cite{Chacon_Tarazona_PhysRevLett} contrary to the
contraction that is found). Nevertheless, the repetition of the
analysis by means of the Smoluchowski Equation (\ref{eq:smoluloss}),
shows no measurable dependence of $D_\perp$
with the area within our $10\%$ error bar.  This supports the
identification of $D_\perp$ as a true intrinsic property of the
surface, and confirms the methods described. Indeed, simpler methods
based on dividing the system in slabs,\cite{buhn_2004, liu} or
selected ``outmost'' molecules\cite{Taylor_1996,taylor_2003} would
have resulted in a clear dependence of $D_\perp$ with the area.

\section{Conclusions and future work}
\label{conclusions}

%% Reworking all this:

We have described an application of the intrinsic sampling method to
the analysis of dynamical processes at the liquid-vapor interface.
The main conclusion is that a liquid surface is a region of enhanced
molecular mobility, with respect to that in the bulk liquid, but
without strong anisotropy.  The characterization of the normal
diffusion constant for the molecules at the outmost liquid layer
requires a much more elaborated method that for the transverse
component, but at the end of the day we get similar values for
$D_\perp$ and $D_\|$, both at high and low temperatures for the simple
liquid models studies in this article. This fact is consistent with
our results for the residence time, which governs turnover rate, i.e.,
the rate at which molecules enter and leave the intrinsic surface ---
a time that turns out to be comparable to the typical times for the
threshold of diffusion. Thus, the typical time required for a molecule
to travel a distance on the order of a molecular diameter is very
similar to its residence time as a surface molecule. Therefore, even if
diffusion coefficients can still be computed for molecules that stay
at the surface for times long enough, the turnover processes are
equally important when discussing the dynamical properties of the
interface. We have discussed two main features of these processes: the
overall residence time, and the input/output rate per molecule,
$\nu_\mathrm{io}(z)$, the spacial distribution function of the
turnover rate.

The particular details of interfacial dynamics will be, of course,
model-dependent. We next provide some relevant cases on which the
method describe here could be applied. References will be given to
previous works on these systems, but we would like to emphasize that
these works have employed the usual approach (by means of slabs),
therefore the present approach (by means of the IS) could shed new
light on the structure and dynamics of systems of considerable applied
interest.

The most important applications of this technique would now be
realistic models of complex fluids, such as the liquid-vapor interface
of water\cite{liu_2005,liu} In this case, the more orderly nature of
liquid phase, and the high surface tension, are likely to provide
additional stability for the molecules at the interface, thereby
leading to longer residence times and more surface molecules reaching
the diffusion regime. We have already started an effort to study this
system from the point of view presented in this article --- in
particular, the discrepancy in our value of the normal diffusion
coefficient for the LJ fluid with results of Ref.~\onlinecite{liu}
suggest that the corresponding values for water will likewise differ.
We also intend to employ this approach to clarify the controversy
surrounding normal diffusion in liquid-liquid interfaces.%
\cite{meyer_1988,benjamin_1992,michael_1998,buhn_2004}

Similar systems of interest include aqueous solutions%
\cite{paul_2005,paul_2005_2,paul_2005_3,chang_2006}
(the later is a useful review on ion solvation at liquid surfaces)
%
%Equally important is the application of this technique to
and surfactants at interfaces.%
\cite{chanda_2005,chanda_2005_2,chanda_2006,clavero_2007} Surfactants
will be usually located at the surface, and most will reach the
diffusive regime.  The relationship between structure and dynamics is
specially interesting in systems with strong profiling, such as liquid
metals,\cite{gonzalez_2006} confined water,%
\cite{martins_2004,sega_2005,marti_2006,hijkoop_2007,lee_1994} and
liquid-solid interfaces.\cite{thomas_2007} Similarly, diffusion in
amphiphilic bilayer structures, \cite{bhide_2005,sega_2005}
micelles,\cite{pal_2005} and microemulsions can be described in a
manner similar to the one presented here.
%

%In this case, it will be interesting to compare the different
%diffusion coefficients (bulk, parallel to the interface, and
%perpendicular to it).

\section{Acknowledgments}

Financial support for this work has been provided by the Direcci\'on
General de Investigaci\'on, Ministerio de Ciencia y Tecnolog\'{\i}a of
Spain, under grants FIS2004-05035-C03, FIS2007-65869-C03, and
CTQ2005-00296/PPQ and Comunidad Aut\'onoma de Madrid under program
MOSSNOHO-CM (S-0505/Esp-0299).

\newpage

\bibliography{biblio}

\end{document}